\documentclass{article}
\usepackage{latexsym}
\usepackage{amsmath}
\usepackage{amssymb}
\usepackage[latin1]{inputenc}   % for german characters ä,ö.ü,ß in *.tex files
\newtheorem{theorem}{Theorem}[section]
\newtheorem{proposition}[theorem]{Proposition} 
 
\newtheorem{lemma}[theorem]{Lemma} 
\newtheorem{example}{Example}[section] 
\newenvironment{proof}{\noindent \bf Proof \rm}{\hspace*{\fill} $\Box $ \vskip7pt}  

\title{Measure and integral with purely ordinal scales}
\author{Dieter Denneberg, Universität Bremen \\ {\small %%@
denneberg@math.uni-bremen.de} \\ \\
Michel Grabisch, Universit\'e de Paris I Panth\'eon-Sorbonne\\ {\small %%@
michel.grabisch@lip6.fr}}
\renewcommand{\labelenumi}{(\roman{enumi})}

\newcommand{\abs}{\mbox{\rm abs}}
\newcommand{\id}{\mbox{\rm id}}
\newcommand{\refl}{\mbox{\rm refl}}
\newcommand{\dist}{\mbox{\rm dist}}
\newcommand{\sign}{\mbox{\rm sign}\,}
\newcommand{\svee}{\,{\bigtriangledown}\,} %{\bigcirc \hspace{-.3 cm} \vee}
\newcommand{\swedge}{\small \bigtriangleup}

\newcommand{\dom}{\mbox{\rm dom}\,}
\newcommand{\graph}{\mbox{\rm graph}\,}
\newcommand{\im}{\mbox{\rm im}\,}
\def\bigsqcap{\mathop{\rule[-0.2ex]{.07em}{2.17ex}
                          \rule[1.8ex]{0.5em}{.17ex}
                          \rule[-0.2ex]{.07em}{2.17ex}}}

\begin{document}

\newcounter{reg1}

\maketitle

%\begin{abstract}
%\end{abstract}

%{\bf Keywords} 

\begin{abstract}
We develop a purely ordinal model for aggregation functionals for lattice
valued functions, comprising as special cases quantiles, the Ky Fan metric and
the Sugeno integral. For modeling findings of psychological experiments like
the reflection effect in decision behaviour under risk or uncertainty, we
introduce reflection lattices. These are complete linear lattices endowed with
an order reversing bijection like the reflection at $0$ on the real interval
$[-1,1]$. Mathematically we investigate the lattice of non-void intervals in a
complete linear lattice, then the class of monotone interval-valued functions and %%@
their inner product.
\end{abstract}

\section{Motivation and survey} 

Measuring and aggregation or integration techniques have a very long tradition. %%@
Here numbers play an important role. But how do humans perceive numbers? The %%@
numbers, say the set $\Bbb R$ of reals, support two basic structures, the %%@
algebraic structure defined by $+$ and $\times$, and the ordinal structure given %%@
by $\le$. There are many situations where only order is relevant, cardinals being %%@
used merely by tradition and convenience. During the last years the interest in %%@
ordinal aggregation has increased, see
e.g.\ \cite{Denneberg 1999, Denneberg and Grabisch 2001, Dubois et al, Grabisch
2000, kos02, mar00a}. 

We are interested in the question if aggregation or integration can be done in
purely ordinal terms and what results can be obtained. Of course many partial
results are already available. Since often they are formulated in terms of
numbers, we ask what can be sustained if one ignores the algebraic structure or
what weaker additional structure has to be imposed on the linear ordinal scale
in order to formulate some well known important issues. It turns out that
enough structure is given by an order reversing bijection of the scale leaving
one point fixed. Thus the scale decomposes into two symmetric parts. This can
be interpreted as the first step to numbers. Since, repeating the procedure
with each resulting part of the scale infinitely often, one ends up with the
binary representation of the numbers in the unit interval $[0,1] \subset {\Bbb
R}$ or some superstructure.

There are several ordinal concepts for aggregating values with respect to a
measure. The oldest and best known selects a certain quantile, say the median, of
a sample as the aggregated value. Next Ky Fan's \cite{Fan} metric on the space
$L_0(\mu)$ of $\mu$-measurable functions is essentially ordinal. More recently
and independently Sugeno \cite{Sugeno 1974} developed his integral which
employs the same idea as Ky Fan. One aim of the present paper is to develop a
common purely ordinal model for these three examples. This is done with a
complete linear lattice $M$ as scale, comprising the classical case $M = [0,1]
\subset {\Bbb R}$. For the Sugeno integral the scales used for functions and
the measure are identical. But in general we allow separate scales for the
functions and the measure and the two scales are related by a commensurability
application, as we call it.

The structure of a linear lattice seems not to be sufficient to model
elementary human behaviour in the presence of risk or uncertainty. There is
some empirical psychological evidence (cf. reflection effect, inverse S-shaped
decision weights, etc.) that in certain decision situations humans have a point
$\Bbb O$ of reference (often the status quo) on their scale which allows to
distinguish good and bad or gains and losses, i.e.\ values above, respectively
below, the reference point \cite{haca94,katv79}. Then the attainable gains and
the attainable losses are aggregated separately and finally these two
aggregated values are compared to reach the final decision. In the cardinal
models this behaviour can be modeled with the symmetric Choquet integral. We
define the symmetric Sugeno integral in order to model the essentials of this
behaviour in purely ordinal terms.

We also define the analogue of the asymmetric Choquet integral in our
context. This can be done in introducing two commensurability functions, one
for the positive part of the scale, the other one for the negative part.

Finally we comment on the new
technical tools and the organisation of the paper.
In Section 2 we model the scale with neutral reference point as a complete
linear reflection lattice $R$ corresponding to $[-1,1] \subset {\Bbb R}$. On
$R$ we use the binary relations from \cite{Grabisch 2000} to get operations
corresponding to addition and multiplication in ${\Bbb R}$.

In Section 3 we develop a theory of increasing interval valued functions and
their inverse. Introducing these tools is motivated as follows.
Mathematically, the idea of Fan and Sugeno for the aggregated value of a random
variable is very simple, just take the argument at which the decreasing
distribution function intersects a preselected increasing function, the
identity function in their case. As already the quantiles show, the aggregated
values are intervals rather than points on the linear scale $L$. So we look for
a suitable ordering on the family ${\cal I}_L$ of nonempty intervals in
$L$. The ordering which had been introduced by Topkis (see \cite{Topkis}) on
the family of nonempty sublattices of an arbitrary lattice turns out to be the
right one to handle monotonic functions (Proposition \ref{mon_corr}). This
ordering, restricted to ${\cal I}_L$, is only a partial one, but we show that %%@
${\cal I}_L$ is a completely distributive lattice in our situation. This structure %%@
is needed for an ordinal analogue of the inner product of vectors, which is %%@
introduced in Section 4. This product will, in Section 7, formalise the idea of %%@
Fan and Sugeno in our general context. Still in Section 4 the product provides a %%@
convenient tool to fill the gaps in the domain of the inverse of a monotone %%@
function, we call this saturation.

In close analogy to probability theory we introduce lattice valued measures in
Section 5 and, in Section 6, the distribution function and its saturated
inverse, the quantile function. In Section 7 we define an aggregated value of a
function $f$ as the product of a preselected commensurability function with the
quantile function of $f$. To meet the empirical findings w.r.t.\ a neutral
reference point mentioned above, we generalise our model in Section 8 to
functions having values in a reflection lattice as introduced in Section 1.

In Section 9 we define the ordinal analogue of a metric induced by a Fan-Sugeno
functional and use this to define the ordinal Ky-Fan 'norm' $\| \cdot \|_0$ and
the ordinal supremum-norm $\| \cdot \|_\infty$.

\section{The reflection lattice}

The order structure of the real line allows for the order reversing reflection
at the null. This structure will be generalised in this section to be the range
of the functions to be integrated. That is, we apply isomorphic scales for gains
and for losses.

Throughout this section $L$ will denote a complete distributive lattice with
bottom ${\Bbb O}$ and top ${\Bbb I}$. We endow the set $L_- := \{ -a \mid a \in
L \}$ with the reversed order from $L$, i.e. $-a \le -b$ in $L_-$ iff $a \ge b$
in $L$. The bottom of $L_-$ is now $-{\Bbb I}$ and the top $-{\Bbb O}$.  The
disjoint union $R$ of $L$ and $L_-$, with $-\Bbb O$ identified with $\Bbb O$,
$$
{\Bbb O} = -{\Bbb O} \,,
$$
and setting $a \le b$ for $a \in L_-$, $b \in L_+$, is called a {\bf reflection
lattice} and $\Bbb O$ the {\bf neutral} or {\bf reference point}. If, in
addition, $R$ is totally ordered, we call it a {\bf linear reflection
lattice}. $R$ is again a (totally ordered) complete distributive lattice and it
has bottom $-{\Bbb I}$ and top ${\Bbb I}$. For emphasising the symmetry we
often write $L_+$ for $L \subset R$, hence $L_+ = \{a\in R \mid a \ge {\Bbb O}
\}$, $L_- = \{a\in R \mid a \le {\Bbb O} \}$. On the reflection lattice $R$ we
have the {\bf reflection} at ${\Bbb O}$
$$
\refl: R \rightarrow R \,, \quad x \mapsto -x \,,
$$
where $-(-a) :=a$ for $a \in R$. The reflection reverses the ordering of $R$,
$$
a<b \,\, \mbox{ iff } -a>-b \,,
$$
$$
(-a) \vee (-b) = -(a \wedge b) \,, \quad (-a) \wedge (-b) = -(a \vee b) \,.
$$
We also define the {\bf absolute value} 
$$
\abs: R \rightarrow L_+ \,, \quad x \mapsto |x|:= \left\{ \begin{array}{ll}
x & \mbox{ if } x\ge {\Bbb O} \\
-x & \mbox{ if } x< {\Bbb O}
\end{array}
\right. 
$$
and the {\bf sign function}  
$$
\sign : R \rightarrow R       %\{{-\Bbb I}, {\Bbb O}, {\Bbb I} \}
\,,\quad \sign x=\left\{ \begin{array}{ll}
{\Bbb I} & \mbox{ for } x > {\Bbb O}\\
{\Bbb O} & \mbox{ for } x = {\Bbb O}\\
-{\Bbb I} & \mbox{ for } x < {\Bbb O}
\end{array}
\right. \,.
$$
 
\begin{example}
{\rm The standard example for a linear reflection lattice is ${\Bbb R} \cup
\{-\infty,\infty\}$ with the usual ordering $\le$ and inf, sup as lattice
operations $\wedge$, $\vee$. Here $-{\Bbb I}= -\infty$, ${\Bbb I}= \infty$ and
${\Bbb O}=0$. Any closed, reflection invariant subset $R$ of this reflection
lattice containing $0$ is again a reflection lattice with ${\Bbb I}= \sup_{x
\in L} x \,$.  \hfill $ {\Box} $ }
\end{example}

Two new operations $\svee$ and $\swedge$ are defined on a reflection lattice
$R$ which will play the roles of addition and multiplication for numbers
(introduced in \cite{gra01d} for linear reflection lattices). On
$L_+$ they coincide with the lattice operations $\vee$, $\wedge$.
$$
x \svee y := \left\{ \begin{array}{ll}
x \vee y & \text{ if } x,y\in L_+\\
x\wedge y & \text{ if } x,y\in L_-\\
x & \text{ if } \sign x \neq \sign y,\,|x|>|y|\\
y & \text{ if } \sign x \neq \sign y,\, |x|<|y|\\
\mathbb{O} & \text{ if } \sign x\neq\sign y,\,|x|=|y| \text{ or } |x|,|y|
\text{ are incomparable.}
\end{array}
\right.
$$
\footnote{Quoting Rubos \cite{rubos}: \emph{\ldots the preorder on preferences is 
partial, and individuals stay with the status quo when unable to make a
comparison}, we see that the definition of $\svee$ for incomparable $|x|,|y|$  %%@
matches with experimental evidence as summarised by Rubos.} 
Except for the last case, $x\svee y$ equals the
absolutely larger one of the two elements $x$ and $y$.   

$$
x \swedge y := \left\{ \begin{array}{ll}
|x| \wedge |y| & \mbox{ if } \sign x = \sign y \\
-(|x| \wedge |y|) & \mbox{ else. }
\end{array}
\right. 
$$
The absolute value of $x \swedge y$ equals $|x| \wedge |y|$ and $x \swedge y <
{\Bbb O}$ iff the two elements $x$ and $y$ have opposite signs.

\begin{proposition}
For $a$,$b$,$c$ in a reflection lattice $R=L_+ \cup L_-$ we have  
\begin{enumerate}
\item
$\svee$ and $\swedge$ are commutative operations;
\item 
${\Bbb O}$ is the unique neutral element of $\svee$ in $R$ and ${\Bbb I}$ the
unique neutral element of $\swedge$.
\item 
$-(a \svee b)= (-a) \svee (-b)$ and $-(a \swedge b)= (-a) \swedge b$;
\item 
$\svee$ is associative on $L_+$ and on $L_-$, $\swedge$ is associative on $R$;
\item
if $a,b,c \in L_+$ or $a,b,c \in L_-$ then  
$a \swedge (b \svee c) = (a \swedge b) \svee (a \swedge c)$\,.
\item
$a \le b$ implies $a \svee c \le b \svee c$ and $a \swedge c \le b \swedge c$.
\end{enumerate}
\label{prop_newop}
\end{proposition}
The straightforward proofs are omitted. Only the proof of (vi) is cumbersome %%@
because of the many cases to be checked. $\svee$ is not associative since
${\Bbb I} \svee ({\Bbb I} \svee (-{\Bbb I}))$ $= {\Bbb I} \neq {\Bbb O} =
({\Bbb I} \svee {\Bbb I}) \svee (-{\Bbb I})$. This happens already in the
smallest non-trivial reflection lattice which is a special case of Example 2.1.

\begin{example}
{\rm In $R =\{-1, 0, 1\} \subset {\Bbb R}$ the operation $\swedge$ coincides
with multiplication in $\Bbb R$ (which is associative) and the operation
$\svee$ is similar to addition with the exception that sums greater than $1$
are pulled down to $1$ and sums below $-1$ are replaced by $-1$.  \hfill $
{\Box} $ }
\end{example}
Non-associativity of $\svee$ will not bother us in the present paper, since we
have only to combine two elements with $\svee$. For a detailed study of
associativity in a linear reflection lattice see \cite{gra01d}.

$\svee$ playing the role of addition, we call $a \svee (-b)$ the {\bf (pseudo)
difference} of $a$ and $b$ in $R$. Then we define the {\bf distance} of $a$ and
$b$ in $R$ by
$$
\dist(x,y) := | x \svee (-y) | = \left\{ \begin{array}{ll}
{\Bbb O} & \mbox{ if } x = y \\
|x| \vee |y| & \mbox{ else }
\end{array}
\right. \,.
$$ 
This distance has the usual properties if $+$ is replaced by $\vee$
\begin{eqnarray*}
\dist(x,y)& \ge & {\Bbb O} \quad \mbox{ and } \dist(x,y)={\Bbb O} \mbox{ iff } x=y %%@
\,, \\
\dist(x,y) & = & \dist(y,x) \,, \\
\dist(x,z) & \le & \dist(x,y) \vee \dist(y,z)\,.
\end{eqnarray*} 
Also $\dist(x,\Bbb O) = |x|$.

\section{Monotone interval-valued functions}

A weakly monotone function is not invertible, but pseudo-inverses are often
used in probability theory. Likewise they are important in our purely ordinal
approach. As the pseudo inverse of a monotone function can be perceived as an
interval-valued function, we use the terminology of set-valued mappings, also
called correspondences. We give two equivalent definitions of isotonicity. For
one of them an ordering for intervals is introduced.

Let $M$ and $L$ be arbitrary sets.  A {\bf correspondence} $\varphi$ from $M$
to $L$ is a mapping $M \rightarrow 2^L$ which assigns to $x \in M$ a subset
$\varphi(x) \subset L$. An ordinary function is a correspondence with
$\varphi(x)$ being a singleton for all $x \in M$. Since $\varphi(x)$ may be
empty for a correspondence $\varphi$, the {\bf domain} of $\varphi$ is defined
as
$$ 
\dom(\varphi) := \{x \in M \mid \varphi(x) \neq \emptyset\} \,.
$$
There is a bijection between correspondences $\varphi$ from $M$ to $L$ and
subsets of $M \times L$ assigning to $\varphi$ its {\bf graph},
$$
\graph(\varphi) := \{ (x,y) \in M \times L \mid y \in \varphi(x) \} \,.
$$
The {\bf inverse correspondence} $\varphi^{-1} : L \rightarrow 2^M$ of a
correspondence $\varphi: M \rightarrow 2^L$ is defined by
$$
\varphi^{-1}(y) := \{x \in M \mid (x,y) \in \graph(\varphi) \}, \quad y \in L
\,,
$$
i.e.\ $\varphi$ and $\varphi^{-1}$ have the same graph (modulo the natural
bijection $M \times L \rightarrow L \times M$).  $\varphi$ is called {\bf
surjective} if its {\bf image} $\im(\varphi) := \dom (\varphi^{-1})$ equals
total $L$.

For defining order preserving correspondences we suppose that $M$ and $L$ are
(partially) ordered sets. A subset $I$ of $L$ will be called a {\bf (preference) %%@
interval}\footnote{
This notion is due to \cite{Fishburn}, the usual definition of (closed, open or %%@
semiclosed) intervals, which is common in  order theory, is not sufficient in our %%@
context.} 
iff for any two points $a,b \in I$ the \textbf{closed interval}
$$
[a,b] :=\{x \in L \mid a \le x \le b\}\,
$$
also belongs to $I$.
Notice that $[a,b]=\emptyset$ if $a>b$. The %%@
intersection of intervals is again an interval and, if $L$ is a linearly ordered %%@
lattice, %%@
also the union of two intersecting intervals is an interval. We denote with ${\cal %%@
%%@
I}_L$ the set of all nonempty preference intervals in $L$. 

A correspondence $\varphi$ from $M$ to $L$ is called {\bf isotonic} or {\bf
increasing} if for all $(x_1,y_2)$, $(x_2,y_1) \in \graph(\varphi)$ with $x_1
\le x_2$, $y_1 \le y_2$ the whole rectangle $[x_1,x_2]\times[y_1,y_2]$ is
contained in $\graph(\varphi)$.  If all such rectangles in $\graph(\varphi)$
are degenerate, i.e.\ $x_1=x_2$ or $y_1 = y_2$, then $\varphi$ is called {\bf
sharply increasing}. Clearly an increasing correspondence $\varphi$ from $M$ to
$L$ is interval valued, so it can be perceived as an application
$$\varphi: D \rightarrow {\cal I}_L \quad \mbox{with } D:=\dom(\varphi) \subset
M \,.$$ Our definition is symmetric in both coordinates of $M \times L$, so the
inverse correspondence $\varphi^{-1}$ of an increasing correspondence $\varphi$
is increasing, too. Especially $\varphi^{-1}$ is interval valued,
\begin{eqnarray}
\label{inv_intv}
\varphi^{-1}: E \rightarrow {\cal I}_M \quad \mbox{with } E:=\dom(\varphi^{-1})
= \im (\varphi) \subset L \,.
\end{eqnarray}

If the ordered set $L$ is a lattice our definition can be formulated more
naturally in introducing an ordering on ${\cal I}_L$.  Topkis (see
\cite{Topkis}) deduces from the ordering of $L$ the {\bf relation
$\sqsubseteq$} on $2^L \setminus \{\emptyset\}$ in defining
$$
Y_1 \sqsubseteq Y_2 \quad \mbox{iff} \quad y_1 \wedge y_2 \in Y_1 \mbox{ and } y_1 %%@
\vee y_2 \in Y_2 \mbox{ for all } y_1 \in Y_1 , y_2 \in Y_2 \,.
$$ 
He shows (\cite{Topkis} Lemma 2.4.1) that the relation $\sqsubseteq$ is
transitive and antisymmetric. Furthermore it is reflexive, hence an ordering,
if it is restricted to the set of nonempty sublattices of $L$ (\cite{Topkis}
Theorem 2.4.1). We excluded the empty set from ${\cal I}_L$ in order to get an
order relation on this set, provided $L$ is linearly ordered (recall that an
open interval in a non-linear lattice is no sublattice in general). ${\cal
I}_L$ does not inherit from $L$ the linearity of the ordering. If $L$ has at
least three elements $y_1 < y_2 < y_3$ then the intervals $[y_2,y_2]$ and
$[y_1,y_3]$ are incomparable w.r.t.\ $\sqsubseteq$ in ${\cal I}_L$.
 
\begin{proposition}
\label{mon_corr}
Let $M$ be an ordered set and $L$ a linearly ordered lattice.  $\varphi : M
\rightarrow {\cal I}_L$ is increasing iff $x_1 \le x_2$ implies $\varphi(x_1)
\sqsubseteq \varphi(x_2)$.
\end{proposition}

\begin{proof}
Let us call the last condition $\sqsubseteq$-increasing. First suppose
$\varphi$ is increasing. Let $x_1 \le x_2$ and $y_1 \in \varphi(x_1)$, $y_2 \in
\varphi(x_2)$. We have to show $y_1 \wedge y_2 \in \varphi(x_1)$ and $y_1 \vee
y_2 \in \varphi(x_2)$. This is obvious if $y_1 \le y_2$, hence we may suppose
$y_1 > y_2$ since $L$ is totally ordered. Since $(x_1,y_1)$, $(x_2,y_2) \in
\graph(\varphi)$ and $\varphi$ is increasing we get $[x_1,x_2]\times[y_2,y_1]
\subset \graph(\varphi)$, especially $y_1 \wedge y_2 = y_2 \in \varphi(x_1)$,
$y_1 \vee y_2 =y_1 \in \varphi(x_2)$.

Now suppose that $\varphi$ is $\sqsubseteq$-increasing. Consider $(x_1,y_2)$,
$(x_2,y_1) \in \graph(\varphi)$ with $x_1 \le x_2$, $y_1 \le y_2$ and take
$(x,y)$ in the rectangle $[x_1,x_2]\times[y_1,y_2]$. We have to show $(x,y) \in
\graph(\varphi)$. We know $\varphi(x_1) \sqsubseteq \varphi(x) \sqsubseteq
\varphi(x_2)$ since $\varphi$ is $\sqsubseteq$-increasing. This entails $y_1 =
y_2 \wedge y_1 \in \varphi(x_1)$ and similarly $y_2 \in \varphi(x_2)$. Hence $y
\in \varphi(x_1)$, $\varphi(x_2)$ and, with an element $y_3 \in \varphi(x)$, $y
\vee y_3$, $y_3 \wedge y \in \varphi(x)$, so $y \in \varphi(x)$.
\end {proof}

A function, perceived as a correspondence, is increasing iff it is an
increasing function in the usual sense (for singletons $\sqsubseteq$ coincides
with the ordering $\le$ on $L$). The results on increasing correspondences
dualise in the obvious manner. So we can use freely {\bf decreasing} (or {\bf
antitonic}) correspondences with their respective properties. As usual {\bf
monotone} means increasing or decreasing.

We investigate the lattice structure of the family ${\cal I}_L$.

\begin{proposition}
\label{lat_intv}
Let $(L,\le)$ be a linearly ordered lattice. Then 
\begin{enumerate}
\item 
$({\cal I}_L,\sqsubseteq)$ is a lattice, containing $(L,\le)$ as the sublattice
of singletons;
\item
$\{\Bbb O\}$ is the bottom and $\{\Bbb I\}$ the top of $({\cal
I}_L,\sqsubseteq)$ if $\Bbb O$ is the bottom and $\Bbb I$ the top of $(L,\le)$;
\item
if $(L,\le)$ is complete, so is $({\cal I}_L,\sqsubseteq)$; 
\item
$({\cal I}_L,\sqsubseteq)$ is distributive. It is completely distributive if
$(L,\le)$ is complete.
\end{enumerate}
\end{proposition}
Under the assumptions of Proposition \ref{lat_intv} the lattice operations join
and meet in $({\cal I}_L,\sqsubseteq)$ will be denoted by $\sqcup$ and
$\sqcap$.  The proof of (i) shows
\begin{eqnarray}
\label{sqcup} 
I_1 \sqcup I_2 = \{a_1 \vee a_2 \mid a_i \in I_i, i=1,2 \}\,, \\
\nonumber
I_1 \sqcap I_2 = \{a_1 \wedge a_2 \mid a_i \in I_i, i=1,2 \}\,,
\end{eqnarray}
and similarly for any number of $\sqcup$ or $\sqcap$ (see (\ref{bigsqcup})).
For closed (and similarly for open) intervals these formulas look more
familiar,
$$[a_1,b_1] \sqcup [a_2,b_2] = [a_1 \vee a_2, b_1 \vee b_2]\,, \quad [a_1,b_1]
\sqcap [a_2,b_2] = [a_1 \wedge a_2, b_1 \wedge b_2].$$ Recall for (iv) that a
linearly ordered complete lattice is completely distributive (\cite{Birkhoff}
V.5).  Proposition \ref{lat_intv} can be generalised to arbitrary
(distributive) lattices $L$ if ${\cal I}_L$ is replaced by the set of closed
nonempty intervals (cf.\ \cite{Nickel 1975} 2.2).  \\

\begin{proof}
(i) Given $I_1$, $I_2 \in {\cal I}_L$, we have to show that a least upper bound
(and similarly a greatest lower bound) of $I_1$ and $I_2$ does exist in ${\cal
I}_L$. Our candidate is
$$J:= \{a_1 \vee a_2 \mid a_i \in I_i, i=1,2 \}\,.$$ First $J$ is nonempty
since the $I_i$ are nonempty. Also $J$ is a (preference) interval. To see this we show that
the closed interval defined by any two points of $J$, say $a_1 \vee a_2 < b_1
\vee b_2$, is contained in $J$.  Let $a_1 \vee a_2 < c < b_1 \vee b_2$, then
for some $i$, $c < b_i$ and, for this $i$, also $a_i < c$. Hence $c \in
I_i$. Let $j$ be the complementary index to $i$, i.e.\ $\{j \}=\{1,2\}\setminus
\{i\}$, then we get $c = c \vee a_j \in J$ as desired.

We just verified that $J \in {\cal I}_L$ and have to show now that $J$ is the
least upper bound of $I_1$ and $I_2$, i.e.\ $K \sqsupseteq I_1$, $I_2$,\; $K
\in {\cal I}_L$ imply $K \sqsupseteq J$. For $a_1 \vee a_2 \in J$ with $a_i \in
I_i$ and $c \in K$ we know $(a_1 \vee a_2) \wedge c = (a_1 \wedge c) \vee (a_2
\wedge c) \in J$ and $(a_1 \vee a_2) \vee c = (a_1 \vee c) \vee (a_2 \vee c)
\in K$. These are the conditions for $K \sqsupseteq J$.

Finally $a \mapsto \{a\}=[a,a]$ defines an isomorphism of $(L,\le)$ to the
sublattice of singletons in ${\cal I}_L$.

(ii) is obvious.

(iii) Let $\{I_i \mid i \in N \}$ be an arbitrary subset of ${\cal I}_L$. Since
$L$ is complete we know $\bigvee_{i \in N} a_i \in L$ for $a_i \in I_i$, $i \in
N$. Now one verifies like for (i) that
\begin{eqnarray}
\label{bigsqcup}
\{\bigvee_{i \in N} a_i \mid  a_i \in I_i, i \in N\}
\end{eqnarray} 
is the least upper bound of the $I_i$, $i \in N$, in ${\cal I}_L$. Compared to
(i) we now need complete distributivity of $L$, but this property holds for a
complete linear lattice as we remarked above. The proof for the greatest lower
bound runs dually.

(iv) Denoting the least upper bound (\ref{bigsqcup}) with $\bigsqcup_{i \in N}
I_i$ and also applying the dual notation, we have to show
$$
\bigsqcap_{k\in K} (\bigsqcup_{i \in N_k} I_{i,k}) = \bigsqcup_{\kappa \in {\cal
K}} (\bigsqcap_{k \in K} I_{\kappa(k),k})
$$
(and the dual equation) for $I_i \in {\cal I}_L$ first with a finite family $K$
of finite index sets $N_k$, $k\in K$, then with an arbitrary family of
arbitrary index sets. Here ${\cal K}$ denotes the set of functions $\kappa : K
\rightarrow \bigcup_{k \in K} N_k$ (the union being disjoint) with $\kappa(k)
\in N_k$. Using (\ref{sqcup}) and (\ref{bigsqcup}), this follows from
(complete) distributivity of $L$,
$$
\bigwedge_{k\in K} (\bigvee_{i \in N_k} a_{i,k}) = \bigvee_{\kappa \in {\cal
K}} (\bigwedge_{k\in K} a_{\kappa(k),k}) \quad \mbox{ for } a_{i,k} \in I_{i,k}
\,.$$
\end{proof}

Finally in this section we extend the ordering and lattice operations on ${\cal
I}_L$ to ${\cal I}_L$-valued applications.  For $\varphi, \psi :M \rightarrow
{\cal I}_L$ we define $\varphi \sqsubseteq \psi$ iff $\varphi(x) \sqsubseteq
\psi(x)$ for all $x \in M$. Similarly $\varphi \sqcup \psi$ and $\varphi \sqcap
\psi$ are defined pointwise.

\begin{proposition}
\label{le_inv}
Let $\varphi$, $\psi :M \rightarrow {\cal I}_L$ be decreasing surjective
correspondences from a complete linear lattice $M$ to a complete linear lattice
$L$. Then we have $\varphi^{-1}$, $\psi^{-1} :L \rightarrow {\cal I}_M$, these
correspondences are decreasing and
$$\varphi \sqsubseteq \psi \;\mbox{ iff }\; {\varphi}^{-1} \sqsubseteq %%@
{\psi}^{-1}\,.$$
\end{proposition}

\begin{proof}
Since $\varphi$, $\psi$ are surjective we know from (\ref{inv_intv}) that
$\varphi^{-1}$, $\psi^{-1} :L \rightarrow {\cal I}_M$ are decreasing. Now it is
sufficient to prove the 'only if' part. Contrary to $\varphi^{-1} \sqsubseteq
{\psi}^{-1}$ we assume that there is a point $y \in L$ so that
${\varphi}^{-1}(y) \not\sqsubseteq {\psi}^{-1}(y)$. Then either the intervals
${\varphi}^{-1}(y)$, ${\psi}^{-1}(y)$ are incomparable in $({\cal
I}_M,\sqsubseteq)$ or ${\varphi}^{-1}(y) \sqsupset {\psi}^{-1}(y)$. In both
cases there exists a point $x \in {\varphi}^{-1}(y)$ with $\{x \} \sqsupset
{\psi}^{-1}(y)$ or a point $x \in {\psi}^{-1}(y)$ with $\{x \} \sqsubset
{\varphi}^{-1}(y)$. Both cases can be treated symmetrically, so we take the
first one. Since ${\psi}^{-1}$ is decreasing and $\{x \} \sqsupset
{\psi}^{-1}(y) \neq \emptyset$ we know that $\graph(\psi)$ cannot intersect the
rectangle $[x,{\Bbb I}]\times [y,{\Bbb I}]$. Now, since $(x,y) \in
\graph(\varphi)$, it is impossible that ${\varphi}(x) \sqsubseteq
{\psi}(x)$. But this contradicts $\varphi \sqsubseteq \psi$.
\end{proof}

\section{Inner product of interval-valued functions}

For interval-valued functions we will construct an ordinal analogue to the
inner product of vectors. We investigate its properties mainly for monotone
correspondences, since this will serve us to define the aggregation functionals
in Section 7. The product will here be applied to saturate monotone
correspondences, i.e.\ to fill the gaps in their domain.

Let $L$ be a complete lattice and $M$ a set.  For $\varphi$, $\psi :M
\rightarrow L$ we define the {\bf (inner) product} as
\begin{eqnarray}
\varphi \ast_M \psi := \bigvee_{x \in M} \varphi(x) \wedge \psi(x) \in L \,.
\label{prod}
\end{eqnarray}
We write $\varphi \ast \psi$ if there is no ambiguity about the common domain
$M$ of $\varphi$ and $\psi$. The name 'inner product' originates from the %%@
following observation.  If $L={\Bbb R}$ and $M$ is finite, then the functions %%@
$\varphi$ and $\psi$ become vectors in ${\Bbb R}^{|M|}$ and the product %%@
(\ref{prod}) resembles the inner product of vectors with $\bigvee$ corresponding %%@
to $\sum$ and $\wedge$ to ordinary multiplication of numbers (see Section 2). With %%@
this interpretation the product behaves 'linear' in both factors as is shown by %%@
properties (iv) and (v) below together with (i).

\begin{proposition}
\label{prop_prod}
Let $L$ be a complete lattice and $M$ a set. For
$\varphi,\varphi_1,\varphi_2,\psi : M \rightarrow L$ and $a \in L$
\begin{enumerate}
\item 
$\varphi \ast \psi = \psi \ast \varphi $\,;
\item 
the 'orthogonality' condition $\varphi \ast \psi = {\Bbb O}$ holds iff
$\varphi$ and $\psi$ have disjoint support;
\item
$\varphi_1 \le \varphi_2\;\;$ implies $\;\;\varphi_1 \ast \psi
\le \varphi_2 \ast \psi$\,;
\item 
$(\varphi_1 \vee \varphi_2) \ast \psi = (\varphi_1 \ast \psi) \vee
(\varphi_2 \ast \psi)\;$ if $L$ is distributive;
\item 
$(a \wedge \varphi) \ast \psi = a \wedge (\varphi \ast \psi)\;$ if $L$ is %%@
completely distributive.  
\end{enumerate}
\end{proposition}

\begin{proof}
(i) The binary relation $\ast$ is commutative since $\sqcap$ is. 

(ii) and (iii) are obvious. 

(iv) Using distributivity of $L$ we
get $(\varphi_1 \vee \varphi_2) \ast \psi = \bigvee_{x \in M} (\varphi_1(x)
\vee \varphi_2(x)) \wedge \psi(x) = \bigvee_{x \in M} (\varphi_1(x) \wedge
\psi(x)) \vee (\varphi_2(x) \wedge \psi(x)) = (\varphi_1 \ast \psi) \vee
(\varphi_2 \ast \psi)$ .

(v) $(a \wedge \varphi) \ast \psi = \bigvee_{x \in M} (a \wedge \varphi(x))
\wedge \psi(x) = \bigvee_{x \in M} a \wedge (\varphi(x) \wedge \psi(x)) = a
\wedge (\varphi \ast \psi) \,.$ For the last equality we applied complete
distributivity of $L$.
\end{proof}

Now we confine ourselves to monotone correspondences and suppose that $L$ is a
complete linear lattice and that $M$ is a linearly ordered set with
bottom ${\Bbb O}$ and top ${\Bbb I}$.
Then, by Proposition \ref{lat_intv}, ${\cal I}_L$ has all the properties
required in  %%@
Proposition \ref{prop_prod} for $L$. 

\begin{example}
\label{Einheitsvektor}
{\rm For $a \in M$ let $\epsilon_a$ denote the indicator function ${\Bbb
I}_{[a,{\Bbb I}]}$ of the interval $[a,{\Bbb I}]\subset M$. Then $\epsilon_a
\ast \psi = \psi(a)$ for any decreasing correspondence $\psi : M \rightarrow {\cal 
I}_L$ from $M$ to $L$. Thus, for monotone functions and in the analogy with the %%@
inner product of vectors, $\epsilon_a$ plays the role of the unit vector for %%@
'coordinate' $a$.  \hfill ${\Box}$ }
\end{example}
In general, the domain $D$ of a decreasing correspondence $\psi: M \rightarrow
2^L$ from $M$ to $L$ is a proper subset of $M$, $D \subsetneq M$. We define the
{\bf saturation} $\widetilde{\psi}$ of $\psi$ as
$$\widetilde{\psi}(x) := \epsilon_x \ast_D \psi\,, \quad x \in M , \mbox{ where
} D=\dom(\psi)\,.$$ $\widetilde{\psi}$ has domain $M$, is interval-valued,
i.e.\ $\widetilde{\psi}: M \rightarrow {\cal I}_L$, and, by Example
\ref{Einheitsvektor}, $\widetilde{\psi}|_D = \psi$.

\begin{proposition}
\label{Prop_satinv}
Let $L$ be a complete linear lattice and $M$ a linearly ordered set with top
and bottom.  The saturation $\widetilde{\psi}$ of a decreasing correspondence
$\psi$ from $M$ to $L$ is decreasing, too.
\end{proposition}

\begin{proof}
For $x_1 \le x_2$ we know $\epsilon_{x_1} \ge \epsilon_{x_2}$, hence by
Proposition \ref{prop_prod} (iii) $\widetilde{\psi}(x_1) = \epsilon_{x_1}
\ast_D \psi \sqsupseteq \epsilon_{x_2} \ast_D \psi =
\widetilde{\psi}(x_2)$. Now apply Proposition \ref{mon_corr}.
\end{proof}

\noindent
If $\psi$ is sharply decreasing, then $\widetilde{\psi}$ is not sharply
decreasing, in general. So we define the {\bf sharp saturation} of a decreasing
correspondence $\psi$ as
$$\widehat{\psi}(x) := \bigvee_{y \in \widetilde{\psi}(x)} y \quad \mbox{for }
x \in M \setminus \dom(\psi)$$ and
$\widehat{\psi}(x):=\widetilde{\psi}(x)=\psi(x)$ for $x \in \dom(\psi)$.
Obviously, $\widehat{\psi} \sqsupseteq \widetilde{\psi}$.

\begin{proposition}
\label{Prop_ssat}
Let $L$ be a complete linear lattice and $M$ a linearly ordered set with top
and bottom.
\begin{enumerate}
\item 
The sharp saturation $\widehat{\psi}$ of a (sharply) decreasing correspondence
$\psi$ from $M$ to $L$ is (sharply) decreasing, too.
\item
Given decreasing functions $\varphi$, $\psi : L \rightarrow M$ then $\varphi
\le \psi$ implies $\widehat{\varphi^{-1}} \sqsubseteq \widehat{\psi^{-1}} \,$.
\end{enumerate}
\end{proposition}
Simple examples show that (ii) does not hold for the saturations
$\widetilde{\varphi^{-1}}$, $\widetilde{\psi^{-1}}$. We supposed $\varphi$,
$\psi$ to be functions only for simplicity.

\begin{proof}
(i) Let $x_1 \le x_2$. We have to show $\widehat{\psi}(x_1) \sqsupseteq
\widehat{\psi}(x_2)$ and distinguish two cases. If $x_2 \in \dom(\psi)$ we know
$\widehat{\psi}(x_1) \sqsupseteq \widetilde{\psi}(x_1) \sqsupseteq
\widetilde{\psi}(x_2)=\widehat{\psi}(x_2)$ from Proposition
\ref{Prop_satinv}. In the second case, $x_2 \notin \dom(\psi)$, we are done if
also $x_1 \notin \dom(\psi)$. For the other case it is sufficient to prove $y_1
\ge {\bar y}_2$ for all $y_1 \in \psi(x_1)$ where $\{{\bar y}_2
\}=\widehat{\psi}(x_2)$. Suppose the contrary, i.e.\ $y_1 < {\bar y}_2$ for
some $y_1 \in \psi(x_1)$. Now the definition of $\widetilde{\psi}(x_2)$
together with the expression (\ref{bigsqcup}) for an arbitrary join of
intervals imply the existence of an $y_3 \in \psi(x_3)$ with $x_3 \in \dom
(\psi)$, $x_3 > x_2$ and $y_1 < y_3 \le y_2$. But monotonicity of $\psi$ would
then imply that the rectangle $[x_1,x_3]\times[y_1,y_3]$ is contained in
$\graph(\psi)$, contradicting $x_2 \notin \dom(\psi)$.

If, in addition, $\psi$ is sharply decreasing, i.e.\ $\graph(\psi)$ does not
contain any non-degenerate rectangle, then the same holds for $\widehat{\psi}$
since $\widehat{\psi}(x)$ is a singleton for any $x \notin \dom(\psi)$.

(ii) Setting $D:=\dom(\varphi^{-1})$, $E:=\dom(\psi^{-1})$ we have, for $y \in
M$,
$$
\widetilde{\varphi^{-1}}(y) = \epsilon_{y} \ast_D \varphi^{-1} = \bigsqcup_{y
\le u \in D} \varphi^{-1}(u) \,,\quad \widetilde{\psi^{-1}}(y) = \epsilon_{y}
\ast_E \psi^{-1} = \bigsqcup_{y \le v \in E} \psi^{-1}(v) \,.
$$
For proving $\widehat{\varphi^{-1}}(y) \sqsubseteq \widehat{\psi^{-1}}(y)$ we
have to distinguish the four cases that $D$ or $E$ contains $y$ or not.

If $y \notin D$ and $y \notin E$ the least upper bounds of the intervals
$\widetilde{\varphi^{-1}}(y)$, $\widetilde{\psi^{-1}}(y)$ have to be
compared. It is sufficient for $\widehat{\varphi^{-1}}(y) \leq
\widehat{\psi^{-1}}(y)$ to find for any point $u \in D$, $u \ge y$ and any
point $x \in \varphi^{-1}(u)$ a point $v \in E$, $v \ge y$ such that $x \in
\psi^{-1}(v)$. By assumption $u \in \varphi(x) \le \psi(x)$. Then $v = \psi(x)$
will do the job.

If $y \in D$ and $y \notin E$ the same argument works since
$\widehat{\psi^{-1}}(y) \sqsupseteq \widetilde{\psi^{-1}}(y)$ is still a
singleton.

Next let $y \notin D$ and $y \in E$. Since $\widehat{\varphi^{-1}}(y)$ is a
singleton it is sufficient for $\widehat{\varphi^{-1}}(y) \sqsubseteq
\widehat{\psi^{-1}}(y)$ to show $\widehat{\varphi^{-1}}(y) \le x$ for any $x
\in \widehat{\psi^{-1}}(y) = \psi^{-1}(y)$. Since $(x,y) \notin
\graph(\varphi)$ we know $y = \psi(x) > \varphi(x)$. $\varphi$ being decreasing
its graph cannot intersect the rectangle $]x,{\Bbb I}]\times]\varphi(x),{\Bbb
I}]$. Hence $\widetilde{\varphi^{-1}}(y) \sqsubseteq [{\Bbb O},x]$ and this
implies $\widehat{\varphi^{-1}}(y) \le x$ as requested.

Finally let $y \in D$ and $y \in E$. For proving $\widehat{\varphi^{-1}}(y)
\sqsubseteq \widehat{\psi^{-1}}(y)$ we apply Lemma \ref{l_sqsubs}. Like above
for any $x \in \widehat{\varphi^{-1}}(y) = \varphi^{-1}(y)$ there exists $v \in
E$, $v \ge y$ such that $x \in \psi^{-1}(v) = \widehat{\psi^{-1}}(v)$. Since
$\psi^{-1}$ is decreasing $\widehat{\psi^{-1}}(v) \sqsubseteq
\widehat{\psi^{-1}}(y)$ by (i), hence there exists $x' \ge x$, $x' \in
\widehat{\psi^{-1}}(y)$. According to Lemma \ref{l_sqsubs} it remains to find
for any $x' \in \widehat{\psi^{-1}}(y)$ a point $x \le x'$, $x \in
\widehat{\varphi^{-1}}(y)$. Since $\varphi(x') \le \psi(x') = y$ we may take
$x=x'$ if $\varphi(x') = \psi(x')$. If $\varphi(x') < \psi(x')$ the graph of
$\varphi$ does not intersect the rectangle $]x',{\Bbb
I}]\times]\varphi(x'),{\Bbb I}]$, hence any point $x \in \varphi^{-1}(y) =
\widehat{\varphi^{-1}}(y)$ will do the job.
\end{proof}

\begin{lemma}
\label{l_sqsubs}
Let $L$ be a linearly ordered lattice and $I_1$, $I_2 \in {\cal I}_L$.  $I_1
\sqsubseteq I_2$ holds iff for each $a_1 \in I_1$ there exists $a_2 \in I_2$
such that $a_1 \le a_2$ and for each $b_2 \in I_2$ there exists $b_1 \in I_1$
such that $b_1 \le b_2$.
\end{lemma}

\begin{proof}
First suppose $I_1 \sqsubseteq I_2$. Given $a_1 \in I_1$ take any $a \in
I_2$. Then $a_1 \vee a \in I_2$ so that $a_2 := a_1 \vee a$ has the desired
properties. The other condition derives similarly. For sufficiency we have to
show $y_1 \wedge y_2 \in I_1$ and $y_1 \vee y_2 \in I_2$ for arbitrary $y_1 \in
I_1$, $y_2 \in I_2$. By assumption there exists $b_1 \in I_1$ such that $b_1
\le y_2$. Then $y_1 \wedge b_1 \le y_1 \wedge y_2 \le y_1$ and since $I_1$ is
an interval containing $y_1 \wedge b_1$, $y_1$ we conclude $y_1 \wedge y_2 \in
I_1$. The other condition proves analogously.
\end{proof}

For the sake of completeness we mention that a {\bf dual product} can be
defined as $\varphi \ast' \psi := \bigsqcap_{x \in M} \varphi(x) \sqcup \psi(x)$
and that the two products are related as follows.

\begin{proposition}
\label{prod<=prod'}
Let $M$ be a totally ordered set and $L$ a complete linear lattice.  If
$\varphi: M \rightarrow {\cal I}_L$ is increasing and $\psi: M \rightarrow
{\cal I}_L$ decreasing, then
$$
\varphi \ast \psi \sqsubseteq \varphi \ast' \psi  \,.
$$
\end{proposition}

\begin{proof}
We are done if we show
\begin{equation}
\varphi(x_1) \sqcap \psi(x_1) \sqsubseteq \varphi(x_2) \sqcup \psi(x_2) \quad 
\mbox{ for all } x_1, x_2 \in M.
\label{eq:33}
\end{equation}
First suppose $x_1 \le x_2$. Using that $\varphi$ is increasing we get
$\varphi(x_1) \sqcap \psi(x_1) \sqsubseteq \varphi(x_1) \sqsubseteq
\varphi(x_2) \sqsubseteq \varphi(x_2)\sqcup \psi(x_2) \,$. Similarly for $x_1
\ge x_2$ we get $\varphi(x_1) \sqcap \psi(x_1) \sqsubseteq \psi(x_1)
\sqsubseteq \psi(x_2) \sqsubseteq \varphi(x_2)\sqcup \psi(x_2) \,$. Since $M$
is totally ordered, the inequality (\ref{eq:33}) is proved for all $x_1, x_2
\in M$.
\end{proof}

\section{Lattice-valued measures}

Probability measures are monotone and assume only nonnegative values. We
maintain this view in our ordinal context. In the cardinal theory of monotone
measures there is a hierarchy of important subclasses: supermodular measures,
totally monotone measures (belief functions), lower chain measures, necessity
measures, ($\sigma$-)additive measures and the hierarchy of the respective
dualisations. In the purely ordinal context among these only the chain measures
and necessity (resp.\ possibility) measures survive. A continuity property will
also be defined in our ordinal environment.

Let $M$ denote a complete linear lattice with bottom ${\Bbb O}$ and top ${\Bbb
I}$. $M$ will be the scale of the measure to be defined. Throughout the paper,
$\Omega$ denotes a nonempty set and ${\cal S}\subset 2^{\Omega}$ a family of
subsets containing $\Omega$ and the empty set, $\emptyset, \Omega \in{\cal
S}$. A $M$-valued set function,
$$
\mu : {\cal S} \rightarrow M
$$ 
is called a {\bf measure}, if $\mu(\emptyset)={\Bbb O}$, $\mu(\Omega)={\Bbb I}$
and it is increasing, i.e.\
$$
A \subset B  \quad \mbox{implies} \quad \mu(A) \le \mu(B) \,.
$$ 

\begin{example}
\label{fuzzy_m}
{\rm The cardinal case $M :=[0,1] \subset {\Bbb R}$ has already been studied
extensively in many different contexts. So $\mu$ has many names in this case,
(Choquet) capacity, non-additive or monotone measure, fuzzy measure, just to %%@
mention the most important ones.  \hfill $ {\Box} $ }
\end{example}

The {\bf inner extension} of a measure $\mu$ is 
$$
\mu_* (A) := \bigvee_{{B \in {\cal S}}\atop B \subset A} \mu(B) = \mu \ast_{\cal %%@
S} \zeta(\,\cdot\,,A)\,, \quad A \in
2^\Omega\,.
$$
Here $\zeta$ denotes the zeta-function of the ordered set $(2^\Omega, \subset)$,
$$
\zeta(B,A) := \left\{ \begin{array}{ll}
{\Bbb I} & \mbox{ if } A \supset B\\
{\Bbb O} & \mbox{ else }
\end{array} 
\right.  \,, \quad A \in 2^\Omega \,.
$$
\footnote
{
In analogy with the cardianl theory one could regard $\mu$, extended with value %%@
${\Bbb O}$ from ${\cal S}$ to $2^\Omega$, as an ordinal Möbius transform of %%@
$\mu_*$. It is not unique since also $\mu_*$ is a Möbius transform of %%@
$\mu_*=(\mu_*)_*$. For details see \cite{gra01d}. 
}
The outer extension is defined dually, $\mu^*:=\mu \ast'_{\cal S} %%@
\zeta(A,\,\cdot\,)$ . Since $\mu$ is increasing, so are
$\mu_*$ and $\mu^*$. For any increasing extension $\nu$ of $\mu$ to $2^\Omega$
\begin{equation}
\mu_* \le \nu \le \mu^* .
\label{ext<=}
\end{equation}

A measure $\mu:{\cal S}\rightarrow M$ is called a {\bf lower chain measure}, if
there is a chain w.r.t.\ set inclusion ${\cal K} \subset {\cal S}$ with
$\emptyset,\Omega \in {\cal K}$ such that
$$\mu=(\mu|{\cal K})_{*}| {\cal S} \,.$$
We refer to $\cal K$ as a {\bf (defining) chain} for $\mu$.
For a lower chain measure $\mu$ on ${\cal S}$,
\begin{equation}
\mu(\bigcap_{A \in {\cal A}} A) = \bigwedge_{A \in {\cal A}} \mu(A) 
\label{minitive}
\end{equation}
for all finite set systems ${\cal A} \subset {\cal S}$ such that $\bigcap_{A
\in {\cal A}} A \in {\cal S}$. The straightforward proof can be found in
\cite{Bruening D} or use Proposition \ref{prop_prod} (iii) together with the fact %%@
that $\zeta$ has property (\ref{minitive}) for the second variable.

If property (\ref{minitive}) holds for arbitrary $\cal A \subset {\cal S}$ and
$\cal S$ is closed under arbitrary intersection, then $\mu$ is called {\bf
minitive} or a {\bf necessity measure}. Dually, one defines {\bf upper chain
measures} and {\bf maxitive} or {\bf possibility measures}.

\begin{example}
\label{uK}
{\rm Like in the cardinal theory we call $u_K:=\zeta(K,\,\cdot\,)$ the {\bf %%@
unanimity game} for 'coalition' $K \subset \Omega$. It is a lower chain measure %%@
with defining chain ${\cal K} :=\{\emptyset, K, \Omega \}$ and it is minitive. But %%@
it is not maxitive, in general. Since $u_K(K^c)={\Bbb O}$ we get
$$
\overline{u}_K (A) := (u_K|\{\emptyset,K^c,\Omega\})^*(A) = \left\{ %%@
\begin{array}{ll}
{\Bbb I} & \mbox{ if } A \cap K \neq \emptyset \\
{\Bbb O} & \mbox{ if } A \subset K^c
\end{array} 
\right.  \,, \quad A \in 2^\Omega \,.
$$
and this is an upper chain measure.  The unanimity game $u_{\{\omega\}}$ for
the singleton $K=\{\omega\}$, often called Dirac measure at point $\omega$,
simultaneously is a lower and upper chain measure.  \hfill ${\Box} $}
\end{example}

\begin{proposition}
\label{min_chain}
Any minitive measure is a lower chain measure. 
Also, any maxitive measure is an upper chain measure.
\end{proposition}
\begin{proof}
Let $\cal K$ be the chain consisting of the sets $K_x := \bigcap\, \{B \mid B
\in{\cal S},\, \mu(B)\ge x \}$, $x \in M$ and $\emptyset$, $\Omega$. We have to
prove $(\mu|{\cal K})_{*}| {\cal S} \ge \mu$ since the reversed inequality
holds by (\ref{ext<=}). Let $A \in{\cal S}$ be arbitrary. With $x=\mu(A)$ we
get $K_{\mu(A)} \subset A$ and by minitivity
\begin{eqnarray*}
(\mu|{\cal K})_{*}(A) &=& \bigvee_{{K \in {\cal S}}\atop K \subset A} \mu(K) \\
&\ge& \mu(K_{\mu(A)}) \; = \; \bigwedge_{{B \in {\cal S}}\atop \mu(B) \ge
\mu(A)} \mu(B) \\ &=& \mu(A) \,.
\end{eqnarray*}
The proof for the maxitive measures runs similarly.
\end{proof}
The converse of Proposition \ref{min_chain} does not hold since, in contrast to
a chain measure, a minitive (maxitive) measure has some continuity property: A
minitive (maxitive) measure is continuous w.r.t.\ the lower (upper)
topology of the complete lattices $2^\Omega$ and $M$ (see \cite{Gierz} III
1.2). But, of course, the class of minitive (maxitive) measures coincides with
the class of lower (upper) chain measures if $\Omega$ is finite.

\section{The quantile correspondence of a lattice-valued function}

Here we do the first steps for aggregating lattice valued functions $f$ on
$\Omega$ w.r.t.\ a monotone lattice-valued measure in introducing the
distribution function of $f$ and the saturation of its inverse, the quantile
correspondence. All this is done in close analogy to probability theory.

Let $\mu : 2^\Omega \rightarrow M$ be a measure, $M$ and $L$ linear lattices
and $f: \Omega \rightarrow L$ a function. Like in probability theory the upper
level sets $\{f\geq x\} := \{\omega\in\Omega \mid f(\omega)\geq x\}$ and $\{f >
x\} := \{\omega\in\Omega \mid f(\omega) > x\}$ of $f$ for level $x \in L$ will
play an important role. Clearly the family of all upper level sets of $f$ forms
a chain. We denote it with ${\cal K}_f \subset 2^\Omega$. A family $F \subset
L^\Omega$ of functions is called {\bf comonotonic} if $\bigcup_{f\in F} {\cal
K}_f$ forms again a chain. For other characterisations of comonotonicity see
\cite{Denneberg 1994}.

The {\bf distribution function} 
$G_{\mu,f} : L \rightarrow M$  of $f$ is defined as 
$$G_{\mu,f}(x) := \mu(f \geq x) \,, \quad x \in L\,.$$ Obviously, $G_{\mu,f}$
is a decreasing function.  Since, in general, ${G}_{\mu,f}$ is not surjective,
the domain of $G_{\mu,f}^{-1}$ can be a proper subset of $M$.  We extend it by
means of the sharp saturation $\widehat{G_{\mu,f}^{-1}}$ defined in Section
4.\footnote 
{ 
In many practical applications the scale $L$ is finite. Then one can avoid the %%@
saturation in replacing ${G}_{\mu,f}$ by the following sharply decreasing %%@
interval-valued correspondence, which is already surjective: $x \mapsto %%@
]\mu(f>x),\mu(f \ge x)]$ for $x \in L \setminus \{{\Bbb I}\}$ (here read $]a,a] = %%@
\{a\}$) and ${\Bbb I} \mapsto [{\Bbb O},\mu(f \ge {\Bbb I})]$. Similarly, in the %%@
classical context with $L={\Bbb R}$, $M=[0,1] \subset {\Bbb R}$ and $\mu$ %%@
continuous from above and below, one has to close the intervals above in order %%@
that the distribution correspondence becomes surjective.  
} 
In analogy with probability theory, we define for $p \in M$ the {\bf $p$-quantile} %%@
of $f$ w.r.t.\ $\mu$ as the interval
\begin{eqnarray*}   
{Q}_{\mu,f}(p) := \widehat{G_{\mu,f}^{-1}} (p)\,\in {\cal I}_L \,.
\end{eqnarray*}
The {\bf quantile correspondence} $Q_{\mu,f}$ is sharply decreasing in the
variable $p \in M$ (Proposition \ref{Prop_ssat} (i)).  If $M$ has the
additional structure of a reflection lattice with fixed point $p_0$ then
${Q}_{\mu,f}(p_0)$ is called the {\bf median} of $f$ w.r.t.\ $\mu$ (cf.\
Example \ref{quant} below).

\begin{proposition}
\label{G_bar}
Let $\mu : 2^\Omega \rightarrow M$ be a measure, $M$ and $L$ complete linear
lattices and $f$, $g \in L^\Omega$ functions. Then
\begin{eqnarray*} 
G_{\mu,f \vee g} \ge G_{\mu,f} \vee G_{\mu,g} \,, \quad 
G_{\mu,f \wedge g} \le G_{\mu,f} \wedge G_{\mu,g}  \\
Q_{\mu,f \vee g} \sqsupseteq Q_{\mu,f} \sqcup Q_{\mu,g} \,, \quad 
Q_{\mu,f \wedge g} \sqsubseteq Q_{\mu,f} \sqcap Q_{\mu,g}  
\end{eqnarray*}
and equality holds if $f$, $g$ are comonotonic. If $\mu$ is an upper (lower)
chain measure then equality holds, too, in the formula for $f \vee g$ ($f \wedge %%@
g$, respectively).
\end{proposition}

\begin{proof}
We first get $G_{\mu,f \vee g}(x) \ge G_{\mu,f}(x) \vee
G_{\mu,g}(x)$ applying the monotone measure $\mu$ on the sets
$$\{f \vee g \ge x \} = \{f \ge x \text{ or } g \ge x \} = \{f \ge x \} \cup
\{g \ge x \} \supset \{f \ge x\}\,,\; \{g \ge x\}\,.$$ If $f$, $g$ are
comonotonic we have $\{f \geq x \}\supset \{g \geq x \}$ or the converse so
that we get an equality. With an upper chain measure we can apply the dual of
(\ref{minitive}), so that we get equality, too.  The assertion with $\wedge$
proves similarly.

The result for the distribution correspondences translates to the corresponding
result for the quantile correspondences by means of Proposition \ref{Prop_ssat}
(ii): $G_{\mu,f \vee g} \ge G_{\mu,f}$ implies $Q_{\mu,f \vee g} =
\widehat{G_{\mu,f \vee g}^{-1}} \sqsupseteq \widehat{G_{\mu,f}^{-1}} =
Q_{\mu,f}$ and similarly with $g$ in place of $f$ on the right hand sides. Both
relations together imply the result.
\end{proof}

The partial order on the set of distribution functions or on the set of quantile %%@
correspondences induces a partial order on $L^\Omega$, often called stochastic %%@
dominance. In the next section we investigate extensions of this order to total %%@
orders.

\section{Fan-Sugeno functionals} 

This is the main part of the article. The former results are applied to define
the class of Fan-Sugeno functionals for lattice valued functions w.r.t.\ a
lattice valued monotone measure and to derive the essential properties.

Here $L$ and $M$ are complete linear lattices. Let $\mu: 2^\Omega \rightarrow
M$ be an $M$-valued measure on a set $\Omega$ and $\ell :M \rightarrow L$ an
increasing function. $\ell$ relates the scale of the measure $\mu$ to the scale
of the functions $f \in L^\Omega$, hence we call it the {\bf commensurability
function}.  The interval-valued {\bf Fan-Sugeno functional} ${S}_{\mu,\ell}:
L^\Omega \rightarrow {\cal I}_L$ is defined by means of the inner product
$\ast$ of Section 4 as
$$
{S}_{\mu,\ell}(f) := \ell \ast Q_{\mu,f} \,.
$$
Always ${S}_{\mu,\ell}(f)$ is a nonempty interval. Often a single value is
preferred, then the least upper bound is the right one. We define
$$
\overline{S}_{\mu,\ell}(f) := \bigvee_{x \in S_{\mu,\ell}(f)} x \,.
$$
If $L=M$ and $\ell$ is the identity mapping $\id_M$ on $M$, i.e.\ $\ell(p)=p$,
$p \in M$, then we write ${S}_\mu$ or $\overline{S}_\mu$ for short. The name,
we attribute to these functionals, deduces from the following special cases.

\begin{example}
{\rm Let $M=[0,1]\subset {\Bbb R}$, $L= [0,\infty]$, $\ell(x)=x$ for $x \in M$
and $\mu$ a probability measure on a $\sigma$-algebra. Since we have defined
the Fan-Sugeno functional only for ordinal measures on the total power set
$2^\Omega$ we first extend $\mu$, say to the inner extension $\mu_\ast$. Now,
for two real random variables $f$, $g$ on $\Omega$, the number
$\overline{S}_{\mu_\ast,\ell} (|f-g|)$ is the distance $\|f-g\|_0$ of $f$ and
$g$ in the Ky Fan metric of the space $L_0(\mu)$ of measurable functions
(\cite{Fan}, see also theorems 9.2.2 and 9.2.3 in \cite{Dudley}). Convergence
in probability is convergence in this metric.  \hfill $ {\Box} $ }
\end{example}

\begin{example}
\label{Sugeno_int}
{\rm Let $L=M=[0,1]$, $\mu$ a fuzzy measure on $2^\Omega$ (Example
\ref{fuzzy_m}) and $f: \Omega \rightarrow [0,1]$. Then
$$
S_\mu (f) = \bigsqcup_{x \in [0,1]} \{x\} \sqcap Q_{\mu,f}(x) \,, \quad
\overline{S}_\mu (f) = \bigvee_{x \in [0,1]} x \wedge G_{\mu,f}(x)\,.
$$
The functional $\overline{S}_\mu (f)$ is the Sugeno integral of $f$ w.r.t.\
$\mu$ (\cite{Sugeno 1974}).  \hfill $ {\Box} $ }
\end{example}
Another special case of our general functional is even better and longer known
than the examples above.

\begin{example}
\label{quant}
{\rm Let $p \in M$ and $\ell:= \epsilon_p$ (see Example \ref{Einheitsvektor}),
then ${S}_{\mu,\epsilon_p}(f) = \epsilon_p \ast {Q}_{\mu,f} = {Q}_{\mu,f}(p)$,
the $p$-quantile of $f$. The classical case is $M=[0,1]
\subset {\Bbb R}$, $L={\Bbb R}$, where ${S}_{\mu,\epsilon_p}(f)$ is the
$(1-p)$-quantile of $f$ in the usual terminology. The difference to our present
terminology results from the fact that we are employing the decreasing
distribution function whereas classically one employs the increasing one. Of
special importance is the case $p=\frac{1}{2}$, then ${S}_{\mu,\epsilon_p}(f)$
is the median of $f$, which, in applications of probability theory, is the
second important location parameter after the expected value.  \hfill $ {\Box}
$ }
\end{example}

\renewcommand{\labelenumi}{(\roman{enumi})}
\begin{proposition}
\label{prop FSO}
Let $L$, $M$ be complete linear lattices. Let $\lambda,\mu: 2^\Omega
\rightarrow M$ be $M$-valued measures on a set $\Omega$ and $k,\ell :M
\rightarrow L$ commensurability functions. The Fan-Sugeno functional
has the following properties where $f,g \in L^\Omega$, $a \in L$:
\begin{enumerate}
\item 
$\ell(\mu(A)) = \overline{S}_{\mu,\ell}({\Bbb I}_A)$ for $A
\subset \Omega$, especially $\mu$ can be reconstructed from $\overline{S}_{\mu}$;
\item 
$
f \le g \mbox{ implies } S_{\mu,\ell}(f) \sqsubseteq S_{\mu,\ell}(g) \,;
$
\item 
If $\ell(\Bbb O)=\Bbb O$, then 
$
S_{\mu,\ell}(a \wedge f) = \{a\} \sqcap S_{\mu,\ell}(f) \,;
$ \item 
$
S_{\mu,\ell}(f \vee g) \sqsupseteq S_{\mu,\ell}(f) \sqcup S_{\mu,\ell}(g)
$ \\
and equality holds if $\mu$ is an upper chain measure;
\item
Comonotonic maxitivity: if $f,g$ are comonotonic, then \\
$S_{\mu,\ell}(f \vee g)=S_{\mu,\ell}(f) \sqcup S_{\mu,\ell}(g)\,$;
\item
$\lambda \le \mu$ and $k \le \ell$ imply $S_{\lambda,k}(f) \sqsubseteq %%@
S_{\mu,\ell}(f)\,.$
\end{enumerate}
\end {proposition}

\begin{proof}
(i) Since $Q_{\mu,{\Bbb I}_A}(\mu(A))=]\Bbb O,\Bbb I]$ and $Q_{\mu,{\Bbb
I}_A}(p)={\Bbb O}$ for $p>\mu(A)$ and ${\Bbb I}$ for $p<\mu(A)$ we see that
$\ell(\mu(A))$ is the top of $\ell \ast Q_{\mu,{\Bbb I}_A} = \bigsqcup_{p \in
M} \ell(p) \sqcap Q_{\mu,{\Bbb I}_A}(p)$.

(ii) From $f \le g$ one easily derives $G_{\mu,f} \le G_{\mu,g}$. Then by
Proposition \ref{Prop_ssat} (ii) $Q_{\mu,f} \sqsubseteq Q_{\mu,g}$ and by
Proposition \ref{prop_prod} (iii) the result follows.

(iii) We know $Q_{\mu,a \wedge f} = Q_{\mu, a}\sqcap Q_{\mu,f}$ from Proposition
\ref{G_bar} since $a$ and $f$ are comonotonic. We have $Q_{\mu,a}(p)=a$, except
for $p=\Bbb O$ and $p=\Bbb I$, where $Q_{\mu,a}(\Bbb O)=]a,\Bbb I]$ and
$Q_{\mu,a}(\Bbb I) = [\Bbb O, a]$. Now observe that $\ell(\Bbb O)\sqcap
Q_{\mu,a}(\Bbb O)= \Bbb O = \ell(\Bbb O)\sqcap \{a\}$ since $\ell(\Bbb O)=\Bbb O$, %%@
and $ Q_{\mu,f}(\Bbb
I)\sqcap Q_{\mu,a}(\Bbb I) = Q_{\mu,f}(\Bbb I)\sqcap \{a\}$ since
$Q_{\mu,f}(\Bbb I)$ is an interval with bottom $\Bbb O$. This proves that
$(Q_{\mu, a}\sqcap Q_{\mu,f}) \ast \ell= (a\sqcap Q_{\mu,f})\ast\ell$. We obtain
the desired result by applying Proposition \ref{prop_prod} (v).

(iv) We know $Q_{\mu,f \vee g}\sqsupseteq Q_{\mu,f}\sqcup Q_{\mu,g}$
(Proposition \ref{G_bar}) and this relation is maintained if we $\ast$-multiply
with $\ell$ (Proposition \ref{prop_prod} (iii)). Finally, the result follows by
distributivity of $\ast$ with $\sqcup$ (Proposition \ref{prop_prod} (iv)).

(v) The proof runs like in (iv).

(vi) From $\lambda \le \mu$ one easily derives $G_{\lambda,f} \le
G_{\mu,f}$. Then by Proposition \ref{Prop_ssat} (ii)
$Q_{\lambda,f} \sqsubseteq Q_{\mu,f}$ and, applying Proposition \ref{prop_prod}
(iii) twice, the result follows.
\end{proof}
We mention some further properties. The transformation rule proves like for the
Choquet integral (see \cite{Denneberg 1994}) and the Fan-Sugeno operators are
compatible with increasing transformations of $L$ and $M$.

For better understanding some of these properties we again employ the analogy
of $\vee$ or $\sqcup$ with the sum of real numbers and of $\wedge$ or $\sqcap$
with the product. Properties (iii) and (iv) tell us that the Fan-Sugeno
functional is a 'linear' operator for upper chain measures. So, in the ordinal %%@
context, the upper chain measures play the role of probability measures.

The dual of properties (iv) and (v) for $f \wedge g$ cannot be proved here since
$\ast$ is distributive with $\sqcup$, but not with $\sqcap$. These properties
for $f \wedge g$ can be obtained with the dual functional, defined with the dual
product $\ast'$ (see proposition \ref{prod<=prod'}). In case of the Sugeno
integral (Example \ref{Sugeno_int}) $\overline{S}_{\mu}$ coincides with its dual
(see e.g. \cite{mar00a}). We illustrate this duality by an example.

\begin{example}
\label{dual_fctional}
{\rm
Let $L=M=[0,1]$, $\ell$ the identity mapping and $\mu$, $f$ so that 
$$
G_{\mu,f}(x) = \left\{ \begin{array}{ll}
1  & \mbox{ for } 0 \le x \le .2 \\
.5 & \mbox{ for } .2 < x \le .6 \\
0  & \mbox{ for } .6 < x \le 1 
\end{array} 
\right .
$$
Then $S_\mu (f) = ].2,.5]$, whereas the dual functional had the value
   $[.5,.6]$.  \hfill $ {\Box} $ }
\end{example}

For $\{0,1 \}$-valued measures our functional coincides with the Choquet integral. %%@
We demonstrate this fact for unanimity games and their conjugate in the next %%@
example.

\begin{example}
\label{S_u_K}
{\rm
Let $M=\{{\Bbb O},{\Bbb  I} \}$, $\ell({\Bbb O})={\Bbb O}$, $\ell({\Bbb I})={\Bbb %%@
I}$ then $S_{\mu,\ell} (f) = ({\Bbb O} \sqcap Q_{\mu,f}({\Bbb O})) \sqcup ({\Bbb %%@
I} \sqcap Q_{\mu,f}({\Bbb I}))= Q_{\mu,f}({\Bbb I})$. Especially for $\mu =u_K$ %%@
(Example \ref{uK}) we get $\overline{S}_{u_K,\ell}(f) = \bigvee \{x \mid \{f \ge x %%@
\}\supset K \} = \bigvee \{x \mid \bigwedge_{\omega \in K} f(\omega) \ge x \} = %%@
\bigwedge_{\omega \in K} f(\omega)$\,. Similarly %%@
$\overline{S}_{\overline{u}_K,\ell}(f) =  \bigvee_{\omega \in K} f(\omega)$\,.
  \hfill $ {\Box} $ }
\end{example}

\section{Fan-Sugeno functionals for $R$-valued functions}

Now we consider functions, for which the range is not only a linear lattice but
has the structure of a linear reflection lattice as introduced in Section
2. Like for the Choquet integral there are two ways to extend the functional of
the last section to functions, taking positive and negative values. One way
results in a functional that is symmetric w.r.t.\ the reflection at $\Bbb O$,
the other one will be an asymmetric functional. Both functionals can be applied
at least to situations where the respective Choquet integrals can be applied.

Let $R$ be a linear reflection lattice with positive part $L=L_+$. For a
function $f:\Omega \rightarrow R$ we define the positive part $f^+$ and the
negative part $f^-$ by
$$
f^+(\omega) := f(\omega) \vee {\Bbb O}\,,\quad f^- := (-f)^+ \,
$$
and one easily sees that $f$ is the difference of $f^+$ and $f^-$, 
$$
f = f^+ \svee (-f^-)\,.
$$
By means of this representation we define the symmetric extension of the
Fan-Sugeno functional $S_{\mu,\ell} : L^\Omega \rightarrow {\cal I}_{L_+}$ with
a commensurability function $\ell : M \rightarrow L_+$ to the {\bf symmetric
Fan-Sugeno functional} $S\!S_{\mu,\ell} : R^\Omega \rightarrow {\cal R}$ as the
pseudo-difference
\begin{eqnarray}
\label{symFS}
S\!S_{\mu,\ell}(f) := S_{\mu,\ell}(f^+) \svee (-S_{\mu,\ell}(f^-)) \,.
\end{eqnarray}
Here the pseudo-addition $\svee$ operates on intervals, not points of $R$,
which remains to be defined. Clearly the disjoint union ${\cal R}$ of ${\cal
I}_{L_-}$ and ${\cal I}_{L_+}$ with $\{-{\Bbb O}\} \in {\cal I}_{L_-}$
identified with $\{{\Bbb O}\} \in {\cal I}_{L_+}$ forms a non-linear reflection %%@
lattice so that the pseudo-addition $\svee$ as defined in Section 2 applies to %%@
${\cal R}$. On ${\cal I}_{L_+}$ the operation $\svee$ coincides with $\sqcup$. 
A noticeable property of $S\!S_{\mu,\ell}(f)$ is that its value is $\mathbb{O}$ as %%@
soon as $S_{\mu,\ell}(f^+)$ and $S_{\mu,\ell}(f^-)$ are incomparable intervals in %%@
${\cal I}_{L_+}$ (see footnote 1). 

Using $(-f)^+ = f^-$, $(-f)^-=f^+$ and $a \svee (-b)= -(b \svee (-a))$
(Proposition \ref{prop_newop} (iii)) one easily checks the symmetry property
$$
S\!S_{\mu,\ell}(-f) = - S\!S_{\mu,\ell}(f) \,.
$$

It is easy to check that defining $\overline{S\!S}_{\mu,\ell}$ by replacing in
(\ref{symFS}) $S_{\mu,\ell}$ by $\overline{S}_{\mu,\ell}$, then the symmetry
property still holds. Also properties (i) and (ii) of Proposition \ref{prop FSO} %%@
still hold for ${S\!S}_{\mu,\ell}$ and $\overline{S\!S}_{\mu,\ell}$ (see %%@
Proposition \ref{prop_newop} (vi)). 

Human behaviour with respect to gains and losses seem not to be symmetric (see
e.g.\ \cite{katv79}). So it might be useful to apply different commensurability
functions for the positive and negative parts. Let $k, \ell : M \rightarrow L_+$
be increasing functions. One may generalise
(\ref{symFS}) to $S\!S_{\mu,k,\ell}(f) := S_{\mu,\ell}(f^+) \svee
(-S_{\mu,k}(f^-)) \,.$

Now we need another pair of increasing commensurability functions, $\ell_+ : M %%@
\rightarrow L_+$ and $\ell_- : M \rightarrow L_- $ for defining 
the {\bf asymmetric Fan-Sugeno functional} $A\!S_{\mu,\ell_-,\ell_+} : R^\Omega
\rightarrow {\cal R}$,
$$
A\!S_{\mu,\ell_-,\ell_+}(f) := S_{\mu,\ell_-}(f) \svee S_{\mu,\ell_+}(f) \,. 
$$
Like before we have to check if the operation on the right hand side is well
defined. This is the case since $\ell_-$ has values in $L_-$ so that
$S_{\mu,\ell_-}(f) \in {\cal I}_{L_-}$ and similar for the other term.
Comparing with (\ref{symFS}), here we have the same function in the two %%@
expressions on the right hand side but different commensurability functions. Also %%@
$-\ell_-$ and $\ell_+$ cannot be compared without an additional structure
of their common domain $L_+$ (to be a reflection lattice, for example) since
$-\ell_-$ is decreasing and $\ell_+$ increasing.

The asymmetric Fan-Sugeno functional is asymmetric in the following sense:
$$
A\!S_{\mu,\ell_-,\ell_+}(-f) = - A\!S_{\mu,-\ell_+,-\ell_-}^\ast (f). 
$$
Here the upper $\ast$ denotes the {\bf conjugate}\footnote {If the scale $M$ of
$\mu$ had the additional structure of an order reversing bijection existing on
$M$, then the {\bf conjugate} $\overline{\mu}$ could be defined like for ${\Bbb
R}$-valued monotone measures and we would get $A\!S_{\mu,-\ell_+,-\ell_-}^\ast
= A\!S_{\overline{\mu},\ell_+,\ell_-}$ (see \cite{Denneberg and Grabisch
2001}).  }  {\bf Fan-Sugeno functional}, which is performed with the increasing
distribution function and the decreasing commensurability correspondences
$-\ell_+$, $-\ell_-$. We leave the details for further research. For the moment we %%@
can say that properties (i), (ii) and (vi) of Proposition \ref{prop FSO} still %%@
hold for $A\!S_{\overline{\mu},\ell_+,\ell_-}$.

\section{Ordinal metrics and norms}

In the spirit of the classical Example 7.1 we define the {\bf ordinal}
$(\mu,\ell)$-{\bf distance} of $R$-valued functions $f$, $g$. Since
$\dist(f(\omega),g(\omega)) = | f(\omega) \svee (-g(\omega)) |$ is an
$L$-valued function on $\Omega$, $L$ being the positive part $L_+$ of $R$, we
can define
$$
\dist_{\mu,\ell} (f,g) := \overline{S}_{\mu,\ell}(| f \svee (-g) |)\,, \quad
f,g \in R^\Omega \,.
$$ 
Like at the end of Section 2 the usual properties of a distance hold with a
restriction for the triangle inequality,
\begin{eqnarray*}
\dist_{\mu,\ell} (f,g)& \ge & {\Bbb O} \quad \mbox{ and }\;\; \dist_{\mu,\ell}
(f,g)={\Bbb O} \; \mbox{ if }\; f=g \,, \\ \dist_{\mu,\ell} (f,g) & = &
\dist_{\mu,\ell} (g,f) \,, \\ \dist_{\mu,\ell} (f,h) & \le & \dist_{\mu,\ell}
(f,g) \vee \dist_{\mu,\ell} (g,h) \quad \mbox{ if } \mu \mbox{ is an upper
chain measure}.
\end{eqnarray*} 
Let us prove the triangle inequality. Using the triangle inequality of
$\dist$, we have for all $\omega\in \Omega$:
\[
|f(\omega)\svee(-h(\omega))|\leq |f(\omega)\svee(-g(\omega))|\vee %%@
|g(\omega)\svee(-h(\omega))|
\]
which entails for $x \in L$
\[
\{|f\svee(-h)|\geq x\}\subset \{|f\svee(-g)|\geq x\}\cup\{|g\svee(-h)|\geq x\}.
\]
By monotonicity of $\mu$ and the dual of (\ref{minitive}) we get:
\[
G_{\mu,|f\svee(-h)|} \leq G_{\mu,|f\svee(-g)|}\vee G_{\mu,|g\svee(-h)|}.
\]
Then by Proposition \ref{Prop_ssat} (ii) this relation translates to the
corresponding quantile correspondences and is maintained if we $\ast$-multiply
with $\ell$ (Proposition \ref{prop_prod} (iii)). Finally, the result follows by
distributivity of $\ast$ with $\sqcup$ (Proposition \ref{prop_prod} (iv)).

We define the {\bf ordinal} $(\mu,\ell)$-{\bf norm}
$$\| f \|_{\mu,\ell} := \dist_{\mu,\ell}(f,{\Bbb O}) \,.$$ It has (with
$\wedge$ interpreted as multiplication) the homogeneity property of usual
norms,
$$\| a \wedge f \|_{\mu,\ell} = |a| \wedge \| f \|_{\mu,\ell} \quad \mbox{for }
a \in R \,,$$ which proves with Proposition \ref{prop FSO} (iii). One also
derives the triangle inequality
$$\| f \svee g \|_{\mu,\ell} \le \| f \|_{\mu,\ell} \vee \| g \|_{\mu,\ell}
\quad \mbox{ if } \mu \mbox{ is an upper chain measure}.$$

Important special cases with $L=M$ and $\ell= \id$ are the ordinal Ky-Fan norm
w.r.t.\ $\mu$
$$\| f \|_{\Bbb O} := \| f \|_{\mu,\rm id}$$
and the $\mu$-essential supremum (cf.\ \cite{Denneberg 1994} Chapter 9)
$$\| f \|_\infty := \| f \|_{\rm sign(\mu), {\rm id}} \,.$$
$f$ is called a $\mu$-{\bf nullfunction} if $\| f \|_\infty = {\Bbb O}$.

\end{document}